\documentclass[useAMS]{pasa}
\usepackage[authoryear]{natbib}

\newcommand{\kms}{\,km\,s$^{-1}$}

\newcommand{\lesssim}{{\leavevmode\kern0.3em\raise.3ex\hbox{$<$}
                       \kern-0.8em\lower0.7ex\hbox{$\sim$}\kern0.3em}}

\title[The Baryonic Tully Fisher Relation]{The Baryonic Tully Fisher Relation}
\author[Gurovich]{Sebasti$\acute{a}$n Gurovich,$^{1,2}$  Stacy S.\ McGaugh,$^{3}$  Ken C.\ Freeman, $^{1}$ Helmut Jerjen,$^{1}$ \\ Lister Staveley-Smith, 
$^{2}$ W.\,J.\,G.\ De\,Blok$^{4}$} 
\affil{
$^1$Mount Stromlo Observatory, Australian National University, Weston Creek, ACT 2611, Australia\\
$^2$Australia Telescope National Facility, CSIRO, Marsfield, Australia \\
$^3$Department of Astronomy, University of Maryland, College Park, MD 20742-2421, USA\\
$^4$Department of Physics and Astronomy, Cardiff University, 5 The Parade, Cardiff CF24 3YB }
\begin{document}

\maketitle

\label{firstpage}

\begin{abstract}
We validate the baryonic Tully Fisher (BTF) relation by exploring the Tully Fisher (TF) and BTF properties of optically and HI-selected disk galaxies. The data includes galaxies from: Sakai et al.\ (2000) calibrator sample;
 McGaugh et al.\ (2000: MC2000) I-band sample; and 18 newly acquired HI-selected field dwarf galaxies observed with the ANU 2.3\,m telescope and the ATNF Parkes telescope from Gurovich's thesis sample (2005).

As in MC2000, we re-cast the TF and BTF relations as relationships between baryon mass and $W_{20}$. First we report some numerical errors in MC2000. Then, we calculate weighted bi-variate linear fits to the data, and finally we compare the fits of the intrinsically fainter dwarfs with the brighter galaxies of Sakai et al.\ (2000). With regards to the local calibrator disk galaxies of Sakai et al.\ (2000), our results suggest that the BTF relation is indeed tighter than the TF relation and that the slopes of the BTF relations are statistically flatter than the equivalent TF relations. Further, for the fainter galaxies which include the I-band MCG2000 and HI-selected galaxies of Gurovich's thesis sample, we calculate a break from a simple power law model because of what appears to be real cosmic scatter. Not withstanding this point, the BTF models are marginally better models than the equivalent TF ones with slightly smaller $\chi_{\rm red}^{2}$.
\end{abstract}

\vspace*{-23.5mm}
\section{Introduction}\vspace*{-1.0mm}

McGaugh et al.\ (2000: MC2000) provides a valuable sample of I-band data
for 63 optically selected dwarf galaxies. 
Several of these galaxies are isolated field dwarfs and a few are among
the slowest rotators of the known population of disk galaxies. 
Using these galaxies, supplemented by other brighter galaxies, MC2000
conjected that a log vs.\ log baryonic TF relation (BTF) which includes all
the visible baryons (gas plus stars) appears to be more nearly linear
than the TF relation which includes only the stellar component. If true,
a BTF relation relates the baryon content to the dark matter
halo's of galaxies, and was probably established at the time of
their formation.

Gurovich, Freeman, and McGaugh (in preparation) have shown that arithmetic
errors crept into the MC2000 I-band data. The I-band
magnitudes were systematically over-luminous by $2 \times (V-I)$ (sign error) and the
HI masses were overestimated by 0.11 dex ($H_0$ error). In order to establish
the validity of the BTF relation we went back to the original
published data and reanalysed it following the methodology of MC2000.
The empirical BTF relation (Freeman,\ 1999; MC2000) for disk galaxies
relates the total baryon mass (gas + stars) and $W_{20}$, the 20th
percentile width of the HI profile. For the brighter galaxies $W_{20}$
is a fairly direct measure of the amplitude of the flat part of the
rotation curve, but for the less massive and more slowly rotating galaxies this is not so clear.
\vspace*{-4.2mm}
\section{Calculating Stellar and Gas Baryons}\vspace*{-1.0mm}

We went to the source of the MCG2000 I-band data for the $64$ dwarf galaxies: Pildis et al.\ (1997), Schombert et al.\ (1997) and Eder \& Schombert (2000), and, from apparent magnitudes kindly made available by James Schombert (private communication), we recalculated the stellar and gas baryon masses following the MCG2000 formulation. 
For all $64$ galaxies we measured the velocity widths $W_{20}$ based on the spectra of Eder \& Schombert (2000) and corrected for the inclinations by Pildis et al.\ (1997).
\begin{figure*}
\includegraphics{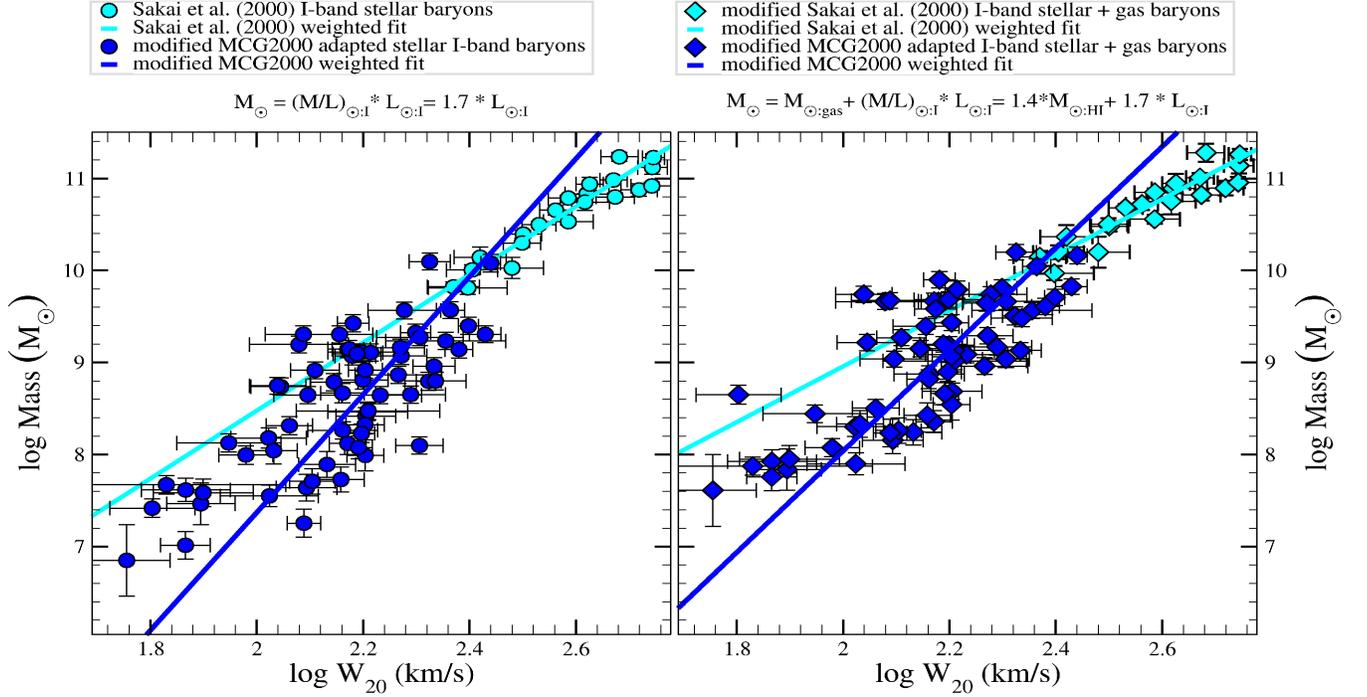}
\vspace*{90mm}
\caption{TF (left panel) and BTF (right panel) plots for the, corrected I-band MC2000 and adapted Sakai et al.\ (2000) I-band calibrator, data
showing the correlation between stellar mass and velocity width
$W_{20}$. Stellar masses are represented by circles. ``Total'' masses, which
include the stellar plus gas components, are represented by diamonds. The stellar and ``total'' mass bi-variate weighted fits are
shown. It is clear that the BTF fits are flatter but, that the TF and BTF relations break for the fainter MCG2000 sample. For the TF and BTF Sakai I-band data the $\chi_{\rm red}^{2}$ are $0.958$ and $1.035$ with slopes of $3.7 \pm 0.3$ and $3.0 \pm 0.3$ respectively and for the TF and BTF MCG2000 I-band data the $\chi_{\rm red}^{2}$ are 4.4 and 4.1 with slopes of $6.4 \pm 0.3$ and $5.5 \pm 0.3$ respectively.}
\label{fig1}
\end{figure*}

\begin{figure*}
\includegraphics{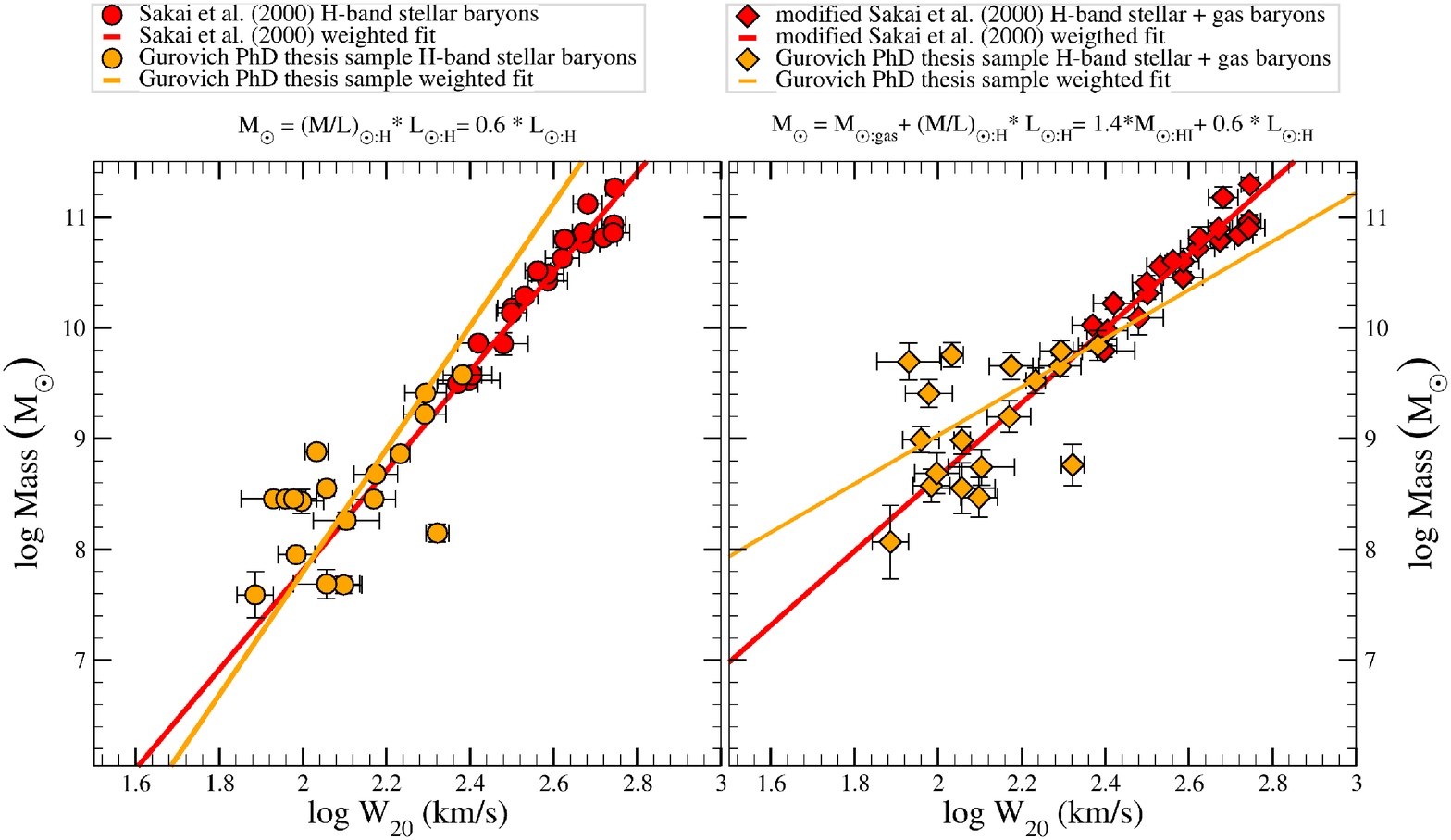}
\vspace*{90mm}
\caption {TF (left panel) and BTF (right panel) plots for the H-band data of the HI-selected galaxies from Gurovich's thesis sample and, for the adapted Sakai
et al.\ (2000) H-band calibrator data. Stellar masses are represented by circles. ``Total'' masses, which
include the stellar plus gas components, are represented by diamonds. The stellar and ``total'' mass bi-variate weighted fits are
shown. It is clear that the slopes of the BTF fits are flatter, but that the TF and BTF relations breakdown for the fainter HI-selected sample. For the HI-selected sample the TF and BTF $\chi_{\rm red}^{2}$ are 9.7 and 6.9 with slopes $5.5 \pm 0.9$ and $2.2 \pm 0.4$ respectively and for the Sakai H-band data the TF and BTF  $\chi_{\rm red}^{2}$ are $0.868$ and $1.109$ with slopes of $4.4 \pm 0.3$ and $3.3 \pm 0.3$ respectively.}
\label{fig2}
\end{figure*}
The MC2000 data come from an optically selected sample of dwarf
galaxies. In this analysis we have also included 18 HI-selected field
dwarf galaxies with inclinations $> 45^\circ$ from the HIPASS survey, 
which are part of a larger
thesis study by Gurovich (2005). In this paper we have also included the brighter local calibrator sample (with Cepheid distances) of Sakai et al.\ (2000). For the HI-selected galaxies we collected H-band
photometry from the Siding Spring 2.3\,m telescope and obtained $W_{20}$
and HI masses from narrow-band (8\,MHz, 1024 channels)
observations with the Parkes telescope. All distances were determined using cosmological redshifts, from
the HI systemic velocities, calculated for $H_0 = 75$\kms Mpc$^{-1}$, except where reliable secondary distances were available. The HI data for the Sakai et al.\ (2000) galaxies were calculated from Martin\ (1998) and Pilyugin et al.\ (2004) but rescaled to the Sakai et al.\ (2000) distances. Total magnitudes are extrapolated to infinity. We performed a TF and BTF analysis on all four data sets by following MC2000 and adopted $M/L_{\rm I} = 1.7$, $M/L_{\rm H} = 0.6$ and estimated the gas mass $M_{\rm g} = 1.4 M_{\rm HI}$ (from cosmological abundances).
\vspace*{-3mm}
\subsection{Weighted Bi-variate Fits}\vspace*{-1.0mm}
We used {\tt fitexy} in {\em Numerical Recipes} (Press et
al.\ 1992) to find weighted bi-variate least-squares linear TF and BTF fits. This algorithm fits the line
that minimizes the scatter in both the $\log (W_{20})$ and the $\log M$
variables, weighting the residuals by the uncertainties
in each quantity.\vspace*{-2mm}

\begin{figure}
\includegraphics{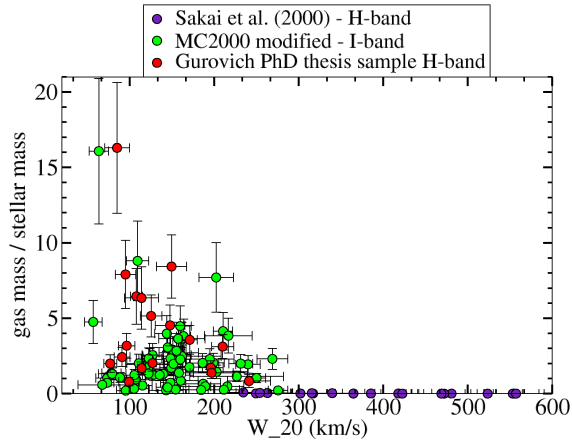}
\vspace*{52mm}
\caption{Gas fraction for the optically selected samples of  MC2000 and Sakai et al.\ (2000) and HI-selected sample of galaxies in this paper.}
\label{fig3}\vspace*{-2.0mm}
\end{figure}

\subsection{Errors Used for Weighting the Bi-variate Fits}\vspace*{-1.0mm}
All estimates of baryon masses and $W_{20}$ have errors arising from uncertainties in the: assumed distances; $W_{20}$ measurements;
integrated HI fluxes (mom0); inclinations and total extrapolated magnitude measurements.

For the Gurovich sample and the corrected MCG2000 I-band data, we adopt random peculiar velocities of 180 \kms\ (2.4 Mpc in distance) from Tonry et al.\ (2000), integrated mom0 errors
of 20\% and total extrapolated magnitude errors of $\sim 0.1$\,mag. We believe that our distance errors are probably overestimates (Karachentsev et al.\ 2003).
The $W_{20}$ errors
are in the range 5\kms$\lesssim\delta W_{20}\lesssim 25$\kms and
8\kms$\lesssim\delta W_{20}\lesssim 50$\kms\ for the two samples, respectively.
The errors for the Sakai et al.\ (2000) galaxies were taken from that paper and the integrated HI flux errors from the standard deviation of the mean values found in the literature re-scaled to the Sakai et al.\ (2000) distances. These errors were propagated following Bevington\ (1969) and subsequently used in the weighted bi-variate fits.

\vspace*{-3mm} \section{Results of the Fits and Conclusions}
The fits are plotted in Figs.\ \ref{fig1} and \ref{fig2}. The slope and $\chi_{\rm red}^2$ can be found in the captions.
These results indicate that for the brighter spiral galaxies (Sakai et al.\ 2000), the BTF fits are tighter than the TF fits and have shallower slopes than the equivalent TF fits. The TF and BTF fits for the fainter galaxies of MCG2000 and of the Gurovich thesis sample break from this simple power law model as is evident from the large $\chi_{\rm red}^2$; $\chi_{\rm red,TF}^2=4.4$ and $\chi_{\rm red,BTF}^2=4.1$ for the I-band
and $\chi_{\rm red,TF}^2=9.7$ and $\chi_{\rm red,BTF}^2=6.9$ for the H-band data, respectively. We would like to highlight the fact that careful consideration went into the error analysis and that this result reveals an inadequate model and is not due to an underestimation of the errors. 
\vspace*{-5.5mm}

\section{Discussion}
\vspace*{-2mm}
Figure 3 shows the gas mass fraction (G/S) for the galaxies in this paper. As is expected, the fainter galaxies are relatively more gas-rich than the brighter spirals of Sakai et al.\ (2000). It is interesting that the brighter spirals which have smaller (G/S) are better modeled by a BTF than a TF model. This is consistent with the conjecture that there is a strict proportionality between the HI and dark Baryons: MCG2000 and Pfenniger and Revaz (2004). It is equally interesting that the fainter disk galaxies that have larger (G/S) are not well modelled by a simple BTF power law fit. It could be that this non-linearity arises partly from the assumptions of the constant $M/L_{I} = 1.7$, $M/L_{H} = 0.6$  and $M_{g} = 1.4 M_{HI}$. In fact Pfenniger and Revaz (2004) conclude that a $M_{g} = 3 M_{HI}$ is a more realistic proportionality.
It is perhaps worth discussing the fainter galaxies in the right panels of Figs. 1 and 2 that appear to be exceptionally baryon-rich for their observed $W_{20}$.

Are these galaxies unusual in having a large baryon mass or a small
value of $W_{20}$?
One possible interpretation is that the gas in these galaxies is not quite at the flat part of the rotation curve and hence their $W_{20}$ values are lower limits of the true rotation amplitude. However, we note that these
galaxies do appear to better follow the (stellar) TF relation.
Another possibility is that the gas fell into these galaxies after
the stellar disk had settled, so that the extra baryons are not part
of the original galaxy.
An even more speculative possibility is that a subsample of these
systems have dark halos which are significantly less
dense than would be expected from the Kormendy \& Freeman (2004)
scaling laws for dark halos, and hence have atypically low values
of $W_{20}$.


 \label{lastpage}

\end{document}